\documentclass[prl,amsmath,amssymb,twocolumn,superscriptaddress]{revtex4-2}

\usepackage{amsmath,amssymb,bm}
\usepackage{graphicx}
\usepackage{xcolor}
\usepackage[colorlinks=true,linkcolor=blue,citecolor=blue]{hyperref}

\graphicspath{{./}{figures/}}

\newcommand{\Pe}{\mathrm{Pe}}
\newcommand{\Ai}{\mathrm{Ai}}
\newcommand{\lb}{\bar{\ell}}
\newcommand{\kc}{\kappa}

\begin{document}

\title{Cutting, Shuffling and Diffusion as a Coarsening Process}

\author{D.~R.~Lester}
\affiliation{School of Engineering, RMIT University, Melbourne, VIC 3000, Australia}
\author{J.~Heyman}
\author{T.~Le~Borgne}
\affiliation{Geosciences Rennes, UMR 6118, Universit\'e de Rennes 1, CNRS, 35042 Rennes, France}

\date{\today}


\begin{abstract}
Cutting and shuffling (CS) rearranges fluid without stretching, but homogenisation requires diffusion (D); this interaction is not well understood. We show that CSD is a coarsening process, with scale $\bar\ell=L(n_c\Pe/2)^{-1/3}$ rather than the Batchelor scale, and that cutting acts as a fractional Laplacian in wavenumber, giving the exact mixing rate $\gamma=2^{1/3}|a_1'|n_c^{2/3}\lambda\Pe^{-1/3}$ for $n_c$ cuts per event at rate $\lambda$, with $a_1'=-1.0188\dots$ the first zero of $\Ai'$. This holds in any dimension for random and deterministic protocols, extends to finite-time
mixing, and places CSD in a known non-equilibrium universality class.
\end{abstract}

\maketitle


Shuffling a deck of cards rearranges without deforming any card. Fluids can be mixed the same way: cut into pieces and reassembled isometrically~\cite{Sturman:2012aa,Christov:2011aa,Smith:2019aa}, forming piecewise isometries (PWIs). This \emph{cutting and shuffling} (CS) arises naturally in discontinuously deforming media such as granular~\cite{Juarez:2010aa,Juarez:2012aa,Christov:2010aa,Zaman:2018aa,Park:2016aa} and yield-stress~\cite{Smith:2017aa,Lester:2019aa} materials, but is also remarkably prevalent in continuous fluids. Indeed, CS can arise in any flow with persistent bifurcations, including valved systems~\cite{Smith:2016aa}, porous~\cite{Lester:2025ae} and fractured~\cite{Ascione:2026aa} media, microfluidic~\cite{Smith:2018aa} and static~\cite{Smith:2019aa} mixers.

Because CS is a measure-preserving isometry, it conserves scalar variance $\sigma^2$ exactly---it stirs but cannot physically mix~\cite{Danckwerts:1952aa}. Note however that CS can mix \emph{ergodically}; PWIs only exhibit weak ergodic mixing~\cite{Katok:1980aa} under specific conditions~\cite{Avila:2007aa}.
Conversely, scalar homogenization involves diffusion (D) acting on the discontinuities that CS generates. Despite its prevalence, a quantitative theory of CSD mixing is lacking. Fig.~\ref{fig:schematic} shows these processes are intertwined: cutting manufactures the gradients diffusion consumes, while diffusion erases the interfaces cutting requires. What emerges from this competition---how fast scalar fields homogenise, and how these are distributed---are the questions we answer here.

To date these have resisted answer. The ergodic theory of the 1D archetype, the interval exchange transformation (IET), is mature~\cite{Keane:1975aa,Katok:1980aa,Avila:2007aa,Novak:2009aa}, and design rules for good finite-time shuffling are established~\cite{Krotter:2012aa,Yu:2016aa,Smith:2018aa}. CS in the presence of diffusion has revealed unexpectedly rich behaviour, including accelerated homogenisation~\cite{Ashwin:2002aa,Sturman:2012aa}, optimal stirring~\cite{Froyland:2016aa}, non-monotonic dependence of mixing rates on diffusivity~\cite{Kreczak:2017aa}, and apparent finite cut-off phenomena~\cite{Wang:2018aa}. These studies characterise mixing via global norms, interface counts and decay rates, with scales estimated by the classical Batchelor scaling $\Pe^{-1/2}$~\cite{Schlick:2013aa,Wang:2018aa}.

Our central claim is that \emph{discontinuous mixing is a coarsening problem}. Each cut creates interfaces; diffusion broadens them; neighbours overlap and merge. The scalar field is therefore a population of diffusing fronts with a distribution of ages, and everything---spectrum, length scale, mixing rate---exactly follows from the balance between front creation and front coalescence. Three physical consequences are worth highlighting:

\begin{figure}[t]
\centering
\begin{tabular}{cc}
\includegraphics[height=0.4\columnwidth]{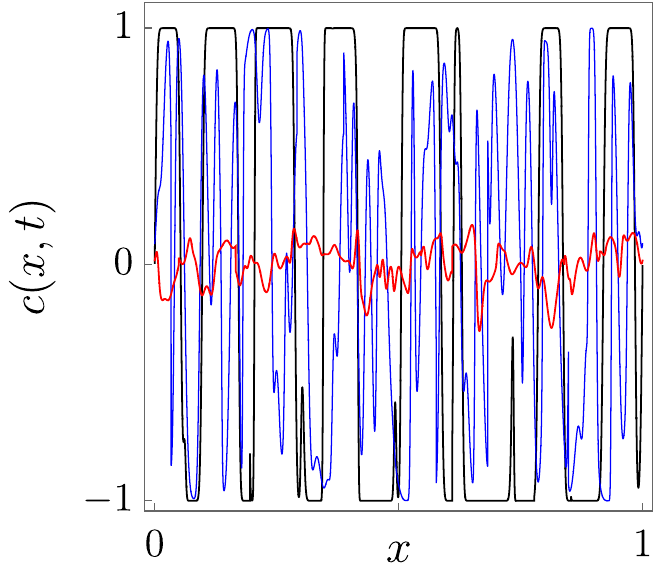}&
\includegraphics[height=0.4\columnwidth]{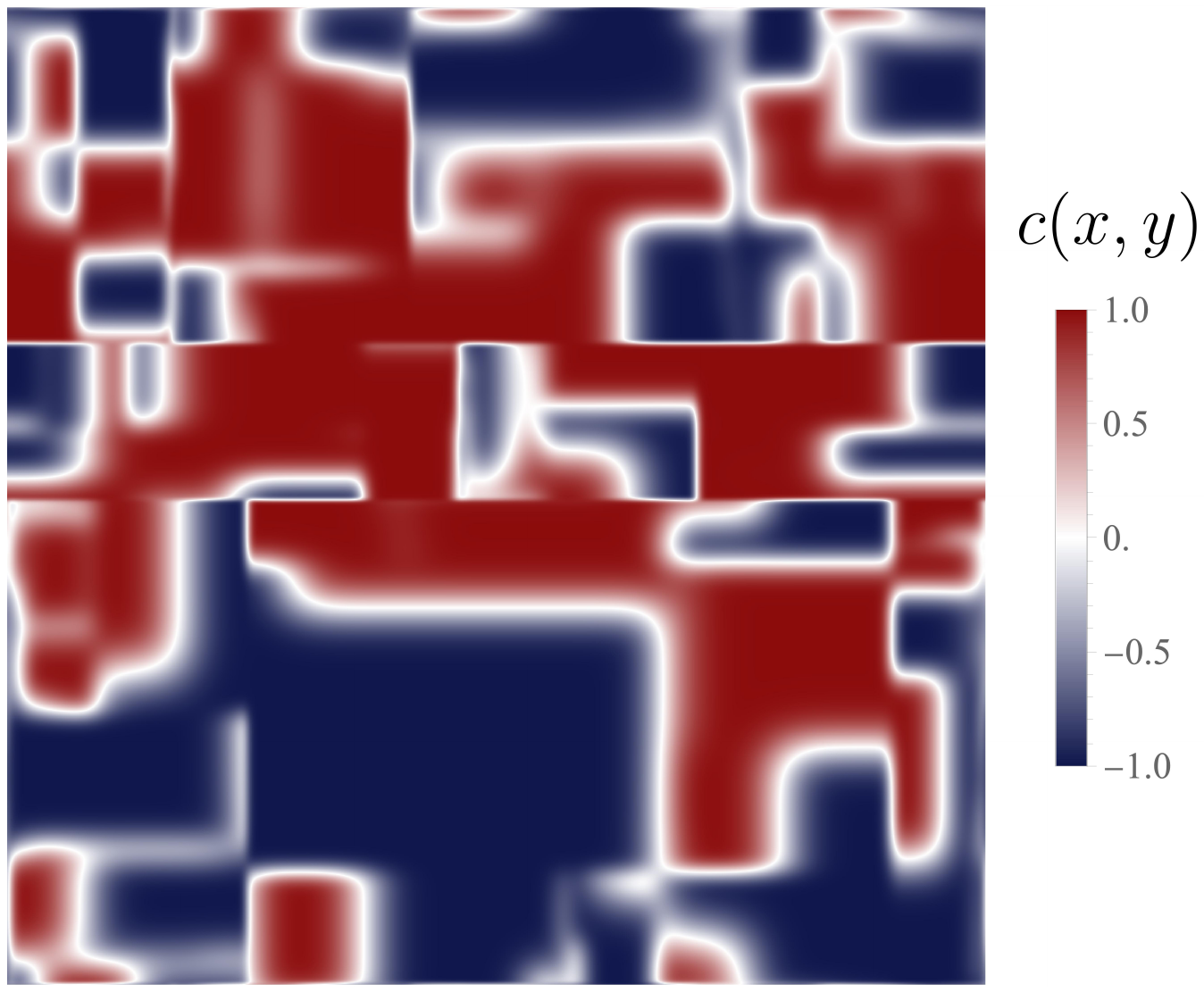}\\
(a) & (b)
\end{tabular}
\caption{Broadening, merger and coarsening of concentration fronts under CSD: (a) transient 1D mixing at $\Pe=10^6$ with $n_c=6$ at $t=4/\lambda$ (black), $6/\lambda$ (blue), $8/\lambda$ (red) and (b) snapshot of 2D mixing at $\Pe=10^4$ via the ``shear'' F1 protocol. Note fronts of various ages (widths).}\label{fig:schematic}
\end{figure}

\emph{(i) The relevant small scale is not the Batchelor scale.} Fronts are created at rate $\kc\equiv n_c\lambda/L$ per unit length, (with $n_c$ cuts per CS event, $\lambda$ the cutting rate, $L$ the domain size) and destroyed once their diffusive width reaches the mean spacing $\lb$, after a time $\sim\lb^2/D_m$. Balancing birth against death gives
\begin{equation}
\lb=L(n_c\Pe)^{-1/3},\qquad \Pe\equiv\lambda L^2/D_m .
\label{eq:lbar}
\end{equation}
This is the balance R\'acz solved for 1D diffusion-controlled annihilation with a steady source~\cite{Racz:1985aa,Krapivsky:2010aa}: thus discontinuous mixing belongs to a known non-equilibrium universality class. The Batchelor-type scale $k_b=\sqrt{\lambda/2D_m}$ only appears as the upper cut-off of a distinct spectral band; they separate as $k_b\lb\sim(\Pe/n_c^2)^{1/6}$, so no single-scale description can survive at large $\Pe$.

\emph{(ii) The spectrum is exact and counts cuts.} With no stretching there is no Batchelor $k^{-1}$ range~\cite{Batchelor:1959aa}; CS is scale-free. Instead, the spectrum is built from discontinuities and their diffusive relics, giving $k^{-2}$ and $k^{-4}$ bands respectively. The amplitude of the terminal $k^{-2}$ band is independent of $D_m$ and equals $4n_c/L^2$: the high-$k$ floor of the spectrum counts the cuts in $n_c$.

\emph{(iii) The mixing rate is exact and algebraic.} CS acts as a fractional Laplacian in wavenumber, giving an Airy eigenvalue problem, with exact mixing rate $\gamma=\dot\sigma^2/\sigma^2$
\begin{equation}
\frac{\gamma}{\lambda}=2^{1/3}|a_1'|\,n_c^{2/3}\,\Pe^{-1/3},
\label{eq:gammaexact}
\end{equation}
where $|a_1'|=1.0188\dots$ is the magnitude of the first zero of $\mathrm{Ai}'$. CS is a genuine mixing mechanism that beats pure diffusion by a factor $\Pe^{2/3}$ but $\gamma/\lambda$ vanishes as $\Pe\to\infty$;  in the absence of stretching, mixing always stays diffusion-limited. This is in sharp contrast with chaotic advection, whose decay rate is set by the stretching rate and is asymptotically $\Pe$-independent~\cite{Pierrehumbert:1994aa,Thiffeault:2008aa}.

\paragraph*{Spectral Theory and DNS.} Before extending to multiple dimensions, we first consider a scalar $c(x,t)$  with diffusivity $D_m$ on the 1D periodic domain $x\in[0,L]$; full derivations are in~\cite{SM}. A CS event is the measure-preserving PWI $\phi$, $c(x,t^+)=c(\phi(x),t^-)$, built from intervals with random or deterministic cut positions and creating $n_c$ cuts, i.e. $n_c$ is twice the number of interval exchanges per CS event. We first consider CS events forming a Poisson (P) process in time at rate $\lambda$, before extending to regular (R) protocols with fixed temporal intervals $\Delta t=1/\lambda$. The theory requires only that successive cuts decorrelate; randomised cut positions achieve this, but so does deterministic and ergodic CS~\cite{SM}. We therefore consider $\phi$ with ergodic cut positions, the exception being fixed IETs, whose resonances dominate earlier studies~\cite{Krotter:2012aa,Kreczak:2017aa}. To generate DNS data we integrate CS with diffusion pseudo-spectrally~\cite{SM} for $\lambda=1$, $L=1$ and
$D_m=10^{-4}$--$10^{-8}$ under Poisson and regular protocols, recording energy spectra after each CS event.

Periodicity confines the spectrum to the wavenumber lattice $k\in(2\pi/L)\mathbb{Z}$ with spacing $\Delta k=2\pi/L$; all sums run over this lattice. Writing $c=\sum_k\hat c(k,t)e^{ikx}$ and $E(k,t)\equiv2\langle|\hat c(k,t)|^2\rangle$ for $k>0$, so that $\sigma^2=\sum_{k>0}E(k,t)$, a CS event acts as $\hat c(k,t^+)=\sum_{k'}A_{kk'}\hat c(k',t^-)$ with $A_{kk'}=L^{-1}\int_0^Le^{-ik\phi^{-1}(x)+ik'x}dx$. Assuming phases decorrelate between events, the covariance $C_{kk'}(t)\equiv\langle \hat{c}(k,t)\hat{c}(k',t)^*\rangle \simeq E_P(k,t)\delta_{kk'}$, with $E_P(k,t)$ the Poisson energy spectrum. For a Poisson CS process of rate $\lambda$ with diffusion, $E_P(k,t)$  evolves as
\begin{equation}
\partial_tE_P=\lambda\!\sum_{k'}\,\!\bigl[T(k,k')-\delta_{kk'}\bigr]E_P(k',t)-2D_mk^2E_P,
\label{eq:master}
\end{equation}
where $T(k,k')\equiv\langle|A_{kk'}|^2\rangle$ represents CS and the last term represents diffusion. The reduction of $C_{kk'}$ to $E_P(k,t)$ is not obvious as IETs are never strongly mixing~\cite{Katok:1980aa}, but it is accurate: measuring $T$ directly from the map and comparing $\sum_{k'}T(k,k')E_P(k',t^-)$ with $E_P(k,t^+)$ gives a sampling error less than 1\% ~\cite{SM}. This closure is exact in the ensemble mean for uniform, independent cut positions, and holds to measurement precision whenever successive cuts decohere~\cite{SM}. As expected, this closure fails  when a cut positions do not decohere, and $\gamma$ falls $2$--$8\times$ below (\ref{eq:gammaexact}) but remains exponential~\cite{SM}.

\begin{figure}[t]
\centering
\begin{tabular}{cc}
\includegraphics[height=0.47\columnwidth]{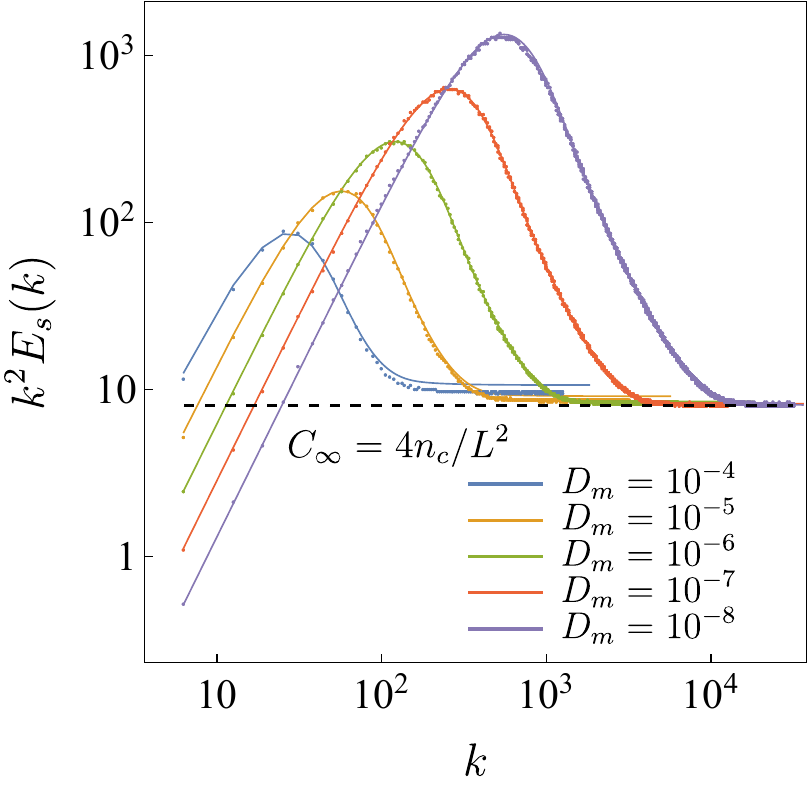}&
\includegraphics[height=0.47\columnwidth]{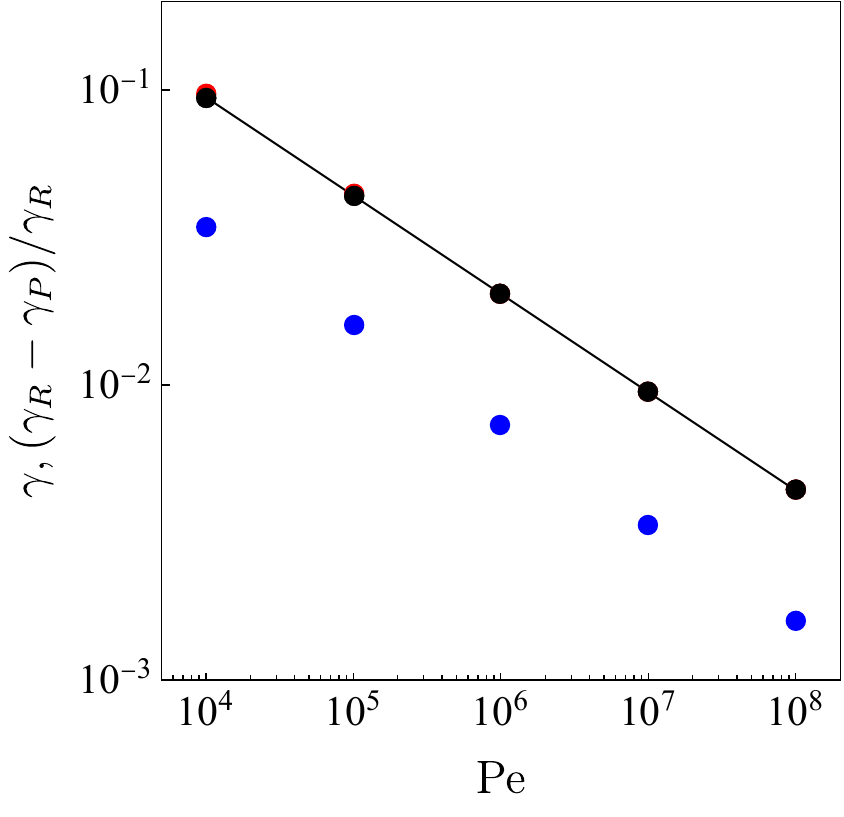}\\
(a) & (b)
\end{tabular}
\caption{(a) The compensated floor $k^2 E_s$ is independent of $D_m$ and counts cuts: $C_\infty=4n_c/L^2$. (b) The Poisson $\gamma_P$ (black dots) and regular $\gamma_R$ (red dots, obscured by black dots) DNS mixing rate matches the exact theory (\ref{eq:gammaexact}) (line) and the relative difference between $\gamma_R$ and $\gamma_P$ scales as $\Pe^{-1/3}$. The branch at $\Pe<10^4$ (not shown) scales diffusively, i.e. $\gamma/\lambda\propto\Pe^{-1}$.}
\label{fig:collapse}
\end{figure}

Conservation of variance makes $A$ unitary, so $\sum_kT(k,k')=1$ and the source amplitude in (\ref{eq:master}) is fixed with no $k$-independent background. Conservation of the mean gives $A_{0k'}=0$, hence $T(k,k')\propto k^2$ as $k\to0$: transfer to large scales is dipolar and no variance reaches the mean. On a periodic domain this constraint acts only below the fundamental mode, and the observable spectrum saturates at a plateau for $k\ll k_c$~\cite{SM}. Writing the asymptotic spectrum as a decaying stationary strange eigenmode~\cite{Ashwin:2002aa} $E_{s,P}(k)$, i.e. $E_P(k,t)\xrightarrow{t}\sigma^2(t)E_{s,P}(k)$ with $\dot\sigma^2=-\gamma\sigma^2$ and mixing rate $\gamma$, Eq.~(\ref{eq:master}) shows $E_{s,P}(k)$ transitions around $k_b=\sqrt{\Lambda_0/2D_m}$, where
\begin{equation}
E_{s,P}(k)=\frac{Q(k)}{\Lambda_0+2D_mk^2},
\label{eq:balance}
\end{equation}
with $Q(k)\equiv\lambda\sum_{k'}T(k,k')E_{s,P}(k')$ and $\Lambda_0=\lambda-\gamma$. Hence $\Lambda_0\rightarrow\lambda$ at large $\Pe$. From (\ref{eq:master}), cutting is a difference kernel~\cite{SM} on the lattice, i.e. CS is a jump process:
\begin{equation}
T(k,k')=\frac{2n_c}{\left[L(k-k')\right]^{2}}, \qquad k\neq k'.
\label{eq:Tiet}
\end{equation}
If the lattice spacing is negligible against the width of the spectrum, i.e.\ $\Delta k\,\lb\ll1$, the sum over $k'$ may be replaced by an integral and $E_P$ is governed by a fractional Laplacian $(-\partial_k^2)^{1/2}$ with a harmonic potential,
\begin{figure}[t]
\includegraphics[width=0.9\columnwidth]{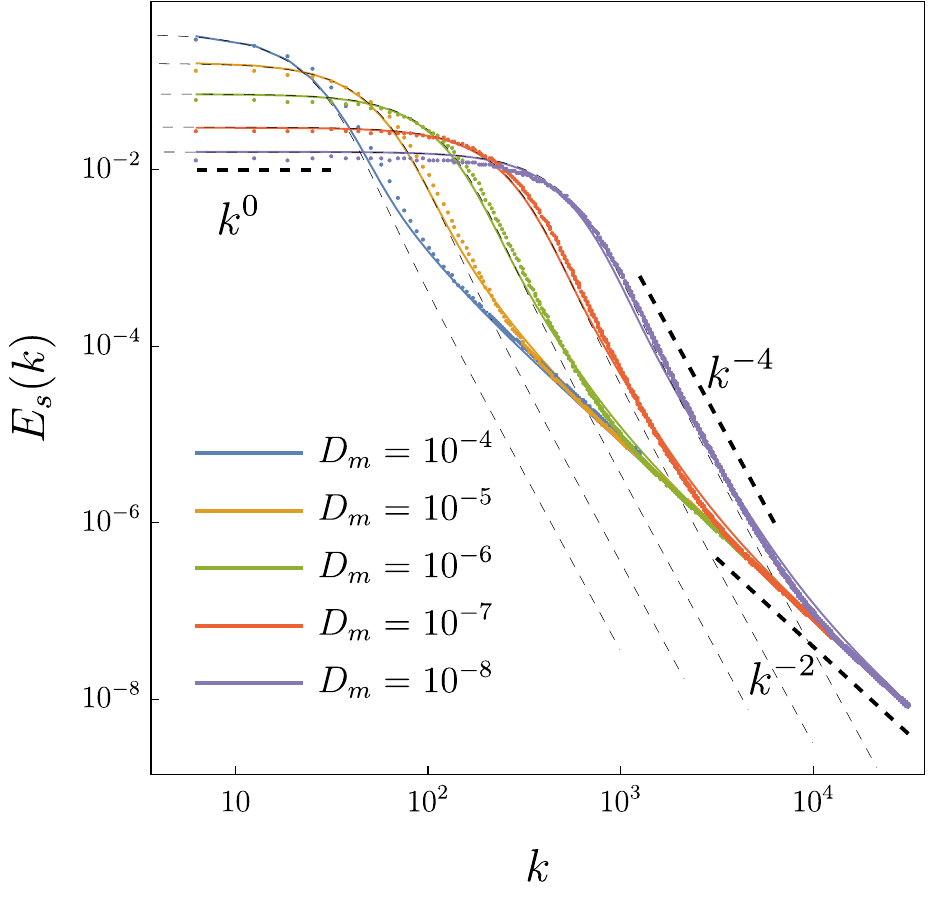}
\caption{Stationary spectra $E_{s,P}(k)$ for Poisson CS from DNS (symbols) and theory Eq.~(\ref{eq:spectrumAi}) (dashed lines) and Eq.~(\ref{eq:regular_corr}) (solid lines). Regular spectra $E_{s,R}^+$ (not shown) are almost identical. Guides show $k^0$, $k^{-2}$ and $k^{-4}$ scaling ranges of Eq.~(\ref{eq:asympt}).}
\label{fig:spectra}
\end{figure}
%
\begin{equation}
\partial_t E_P= -\kc\,(-\partial_k^{2})^{1/2}E_P-2D_mk^2 E_P .
\label{eq:Schrodinger_frac}
\end{equation}
Hence CS is a non-local, heavy-tailed random walk that spreads energy to large $|k|$ while diffusion is a confining potential that removes energy at large $|k|$. Thus (\ref{eq:Schrodinger_frac}) is a continuous-$k$ approximation whose error is quantified below. Because $k$ lives on a lattice of spacing $2\pi/L$, the variable conjugate to $k$ is the physical lag $r$ on the circle $[0,L)$, and the transform of $E_P$ is the autocorrelation $C(r,t)=\sum_{k>0}E_P(k,t)\cos(kr)$. In this variable the lattice multiplier of the cutting term is elementary~\cite{SM},
\begin{equation}
\partial_tC=2D_m\partial_r^2C-\kc\,|r|\left(1-\frac{|r|}{L}\right)C ,
\label{eq:lagexact}
\end{equation}
with Neumann condition $\partial_rC(0,t)=0$. In (\ref{eq:lagexact}), $\kc|r|(1-|r|/L)$ is the rate at which a cut falls between two points separated by $r$, while diffusion acts on the separation with diffusivity $2D_m$. The continuous-$k$ approximation $\psi(r,t)$ of $C$ is precisely the neglect of the curvature term $|r|^2/L$, leaving the linear potential in the Airy equation
\begin{equation}
\partial_t\psi=\mathcal{L}(\psi)\equiv 2D_m\partial^2_{r}\psi-\kc|r|\psi ,\label{eq:Schrodinger}
\end{equation}
which is accurate because the eigenmodes are supported on $r\sim\lb\ll L$. The dominant eigenmode is $\psi_1(r)=\Ai\!\left[(|r|-\gamma/\kc)/\lb\right]$, and the corresponding eigenvalue is the mixing rate $\gamma$ of (\ref{eq:gammaexact}); the full ladder $\gamma_j=(|a_j'|/|a_1'|)\gamma$ follows from the zeros $a_j'$ of $\Ai'$. First-order perturbation theory in the neglected term gives $\gamma/\gamma_{\rm Airy}-1\simeq(M_2/|a_1'|)\,\lb/L$ with $M_2=\int_0^\infty\!s^2\Ai^2/\!\int_0^\infty\!\Ai^2$ evaluated at $s+a_1'$~\cite{SM}, so the continuous-$k$ error is $O(\lb/L)=O(\Pe^{-1/3})$ and vanishes in the mixing limit. A jump in $\psi_1'''(r)$ at $r=0$ produces a $k^{-4}$ band in $E_{s,P}$ as $k\rightarrow\infty$~\cite{SM}, and the full stationary spectrum is
\begin{equation}
E_{s,P}(k)=\frac{4}{L\,\Ai(-|a'_1|)}\int_0^\infty \cos(kr)\,\Ai\!\left[\frac{r}{\lb}-|a'_1|\right]dr,
\label{eq:spectrumAi}
\end{equation}
which is normalised, $\sum_{k>0}E_{s,P}(k)=1$, with $\lb$ from (\ref{eq:lbar}). As shown by Fig.~\ref{fig:collapse}a, its high-$k$ floor (\ref{eq:asympt}) is set by  $n_c$~\cite{SM},
\begin{equation}
E_s(k)\to\frac{C_\infty}{k^2},\quad C_\infty=\frac{4n_c}{L^2}.
\label{eq:Cinfty}
\end{equation}
The transient spectrum $E_P(k,t)$ is also given by the complete eigenmodes of $\mathcal{L}$. Note that closed form approximations for (\ref{eq:spectrumAi}) have also been developed~\cite{SM}. Fig.~\ref{fig:spectra} shows that (\ref{eq:spectrumAi}) captures the DNS data almost exactly except for the $k^{-2}$ tail, which can be explained by consideration of regular CS events as follows.
\paragraph*{Regular CS.} The Poisson solution (\ref{eq:spectrumAi}) can be 
extended to regular CS events with fixed spacing $\Delta t=1/\lambda$. Denoting $E_{R}^+(k,t)$, $E_{R}^-(k,t)$ respectively as the regular spectrum immediately after and before a CS event, then
\begin{equation}
E_R^-(k,t+\Delta t)=e^{-z} E_R^+(k,t),\,\, z\equiv \frac{2D_mk^2}{\lambda},
\label{eq:ER+-}
\end{equation}
and the CS transfer operator $\mathcal{K}$ updates $E_R^+$ as $E_R^+(k,t+\Delta t)=\mathcal{K}e^{-z}E_R^+(k,t)$. This finite-step can be connected to (\ref{eq:spectrumAi}) by consideration of a new discontinuity which produces a large-$k$ spectrum $P(k)\sim k^{-2}$, which then ages with time via diffusive attenuation as $\exp(-z \lambda t)$. For continuous and discrete cutting events, the dimensionless residence factors are respectively 
\begin{equation}
\lambda\int_0^\infty e^{-z\lambda t}\,dt=\frac{1}{z}, \quad \sum_{m=0}^\infty e^{-mz}=\frac{1}{1-e^{-z}},
\end{equation}
hence the finite-step spectrum is given as 
\begin{equation}
E_R^+(k)\approx E_P(k)\frac{z}{1-e^{-z}},
\label{eq:regular_corr}
\end{equation}
and the mixing rate under the regular CS protocol is also slightly larger, in \cite{SM} we show
\begin{equation}
\gamma_R=\gamma_P[1+O(\Pe^{-1/3})],
\label{eq:gamma_RP}
\end{equation}
as confirmed in Fig.~\ref{fig:collapse}b,
%
%
hence $\gamma_R\rightarrow\gamma_P$ as $D_m\rightarrow 0 $. Note that if results are sampled after the diffusion step, then from (\ref{eq:ER+-}), the above spectrum is multiplied by $e^{-z}$. As $z/(1-e^{-z})=1+z/2+O(z^2)$, the continuous spectrum is unchanged for $z\ll1$ and for $z\gg 1$, $E_R^+\sim k^{-2}$, i.e. the finite-step spectrum exhibits a $k^{-2}$ tail. As $d\ln E_R^+/d \ln k\approx -2-2z/(e^z-1)$, the midpoint of the transition (slope $=-3$) is given by $z_b=1.256$. Hence the transition $k_b$ and coarsening $k_c$ crossovers occur at
\begin{equation}
k_b=\left(\frac{z_b \lambda}{2 D_m}\right)^{1/2},\quad k_c=\frac{1}{\lb}=\left(\frac{\kc}{2D_m}\right)^{1/3},
\end{equation}
which separate as $k_b/k_c=(z_b^3\Pe/2n_c^2)^{1/6}$. In \cite{SM} we show that these results are consistent with a physically-motivated model based on diffusive fronts generated by CS events that coarsen with age. This model shows that the fronts generated by regular CS form two distinct populations; fresh discontinuities ($k^{-2}$) and aged fronts ($k^{-4}$), leading to the overall spectrum 
\begin{equation}
E_s(k)\simeq\begin{cases}
6.042\,\lb/L, & k\ll k_c,\\[0.6ex]
2n_c\lambda k^{-4}/D_mL^2, & k_c\ll k\ll k_b,\\[0.6ex]
C_\infty k^{-2}, & k\gg k_b ,
\end{cases}
\label{eq:asympt}
\end{equation}
The $k^{-4}$ band corresponds to the fraction of fronts still sharp at scale $1/k$, which falls as $(k_c/k)^2$ and multiplies the single-front $k^{-2}$. The terminal $k^{-2}$ band is carried by the discontinuities of the most recent cut. Matching the two gives $k^2=\lambda/2D_m$, recovering exactly the $k_b$ of the independent balance (\ref{eq:balance}).

Fig.~\ref{fig:spectra} shows that theory (\ref{eq:regular_corr}) closely matches the Poisson spectra ($<1\%$), and the regular spectra (not shown) are almost identical. This is a sampling-protocol artefact; the Poisson spectra are measured immediately following a CS event, and so the $k^{-2}$ tail of the fresh front dominates, whereas the specific timing of older ages fronts plays a minor role. This is confirmed by Fig.~\ref{fig:collapse}a which shows that the Poisson spectra and theory (\ref{eq:regular_corr}) yield an invariant $k^{-2}$ tail that is independent of $D_m$. Similarly, Fig.~\ref{fig:collapse}b shows the mixing rate $\gamma$ follows (\ref{eq:gammaexact}) closely and both the Poisson $\gamma_P$ and regular $\gamma_R$ rates are very similar, with the relative difference scaling as $\Pe^{-1/3}$.

\begin{figure}[t]
\centering
\begin{tabular}{ccc}
\includegraphics[height=0.33\columnwidth]{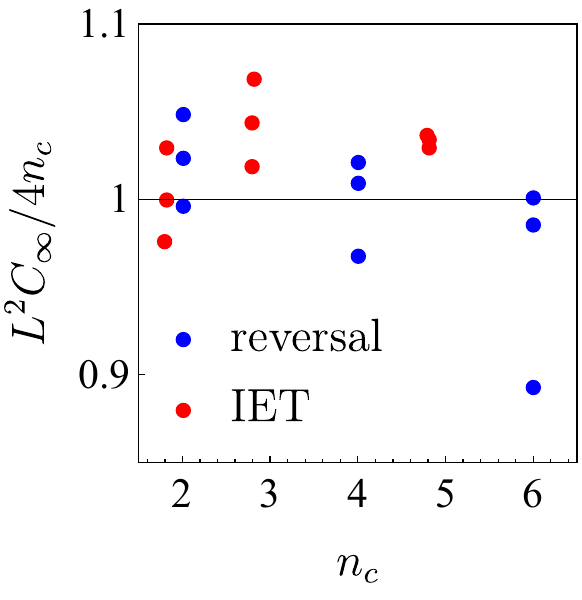}&
\includegraphics[height=0.32\columnwidth]{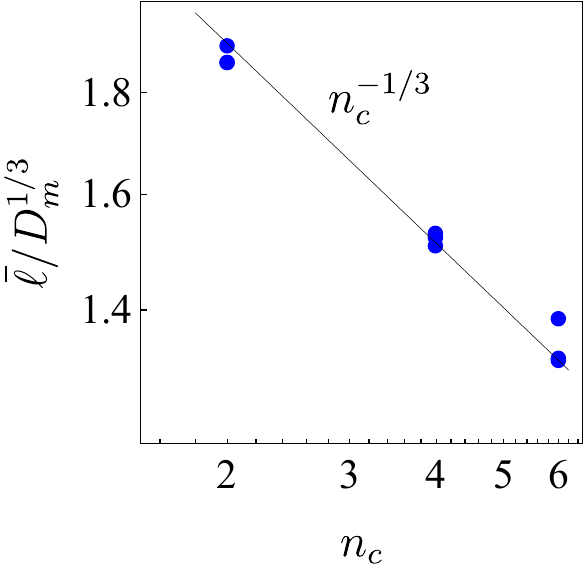}&
\includegraphics[height=0.32\columnwidth]{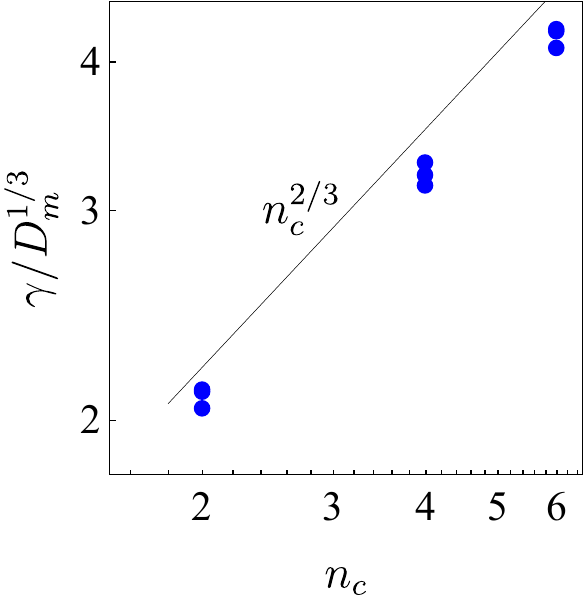}\\
(a) & (b) & (c)
\end{tabular}
\caption{Scaling of various quantities with $n_c$. (a) $L^2C_\infty/4n_c$, Eq. (\ref{eq:Cinfty}),  (thin line), for sub-arc reversals and IETs. (b) $\lb\propto n_c^{-1/3}$, Eq. (\ref{eq:lbar}). (c) $\gamma\propto n_c^{2/3}$, Eq. (\ref{eq:gammaexact}).}
\label{fig:ncsweep}
\end{figure}

The theory predicts three different scalings for the cut number $n_c$: $C_\infty\propto n_c$, $\lb\propto n_c^{-1/3}$  and $\gamma\propto n_c^{2/3}$. Figure~\ref{fig:ncsweep} tests these scalings across two CS protocol families (sub-arc reversals and $m$-interval IETs) and over three values of $\Pe$. We find the DNS results agree with these predicted exponents to within $1\%$ rms error.

\paragraph*{Finite-time mixing.} In general, the finite-time spectrum $E_P(k,t)$ is given by all eigenmodes of $\mathcal{L}$~\cite{SM}
\begin{equation}
E_P(k,t)=\sum_{j=1}^\infty \frac{b_j F_j}{k_A}\left(\frac{k}{k_A}\right)\exp\left[-2^{1/3}n_c^{2/3}|a'_j|\lambda \Pe^{-1/3}t\right],
\label{eq:spectrum_finite}
\end{equation}
with $k_A=1/\lb$, $\Ai'(-|a'_j|)=0$ and $b_j$ known~\cite{SM} for the initial condition $c(x,0)=\mathbf{1}_{0<x<1/2}$. Hence each Airy mode decays as $\gamma_j=2^{1/3}n_c^{2/3}|a'_j|\lambda \Pe^{-1/3}$, and the transient mixing rate evolves as
\begin{equation}
\gamma(t)=2^{1/3}n_c^{2/3}\lambda\,\Pe^{-1/3}\frac{\sum_j |a'_j| b_j \Ai(-|a'_j|)e^{-|a'_j|\tau}}{\sum_j b_j \Ai(-|a'_j|)e^{-|a'_j|\tau}}\xrightarrow{t} \gamma_P,
\end{equation}
with $\tau=t/t_c$, which relaxes to the dominant mode as $\exp(-\Delta_a\tau)$, $\Delta_a=|a'_2|-|a'_1|=2.229$, where the coarsening time is
\begin{equation}
t_c=\frac{\lb^2}{2D_m}=\left(\frac{L}{n_c\lambda}\right)^{2/3}(2D_m)^{-1/3}=\frac{|a'_1|}{\gamma_P}.
\label{eq:tc}
\end{equation}
Extension to regular CS events requires the discrete propagator $E_R^+(k,t=N/\lambda)=(\mathcal{K}e^{-z})^NE_R^+(k,0)$ rather than the asymptotic correction (\ref{eq:regular_corr}).

\begin{figure}[t]
\centering
\begin{tabular}{cc}
\includegraphics[height=0.48\columnwidth]{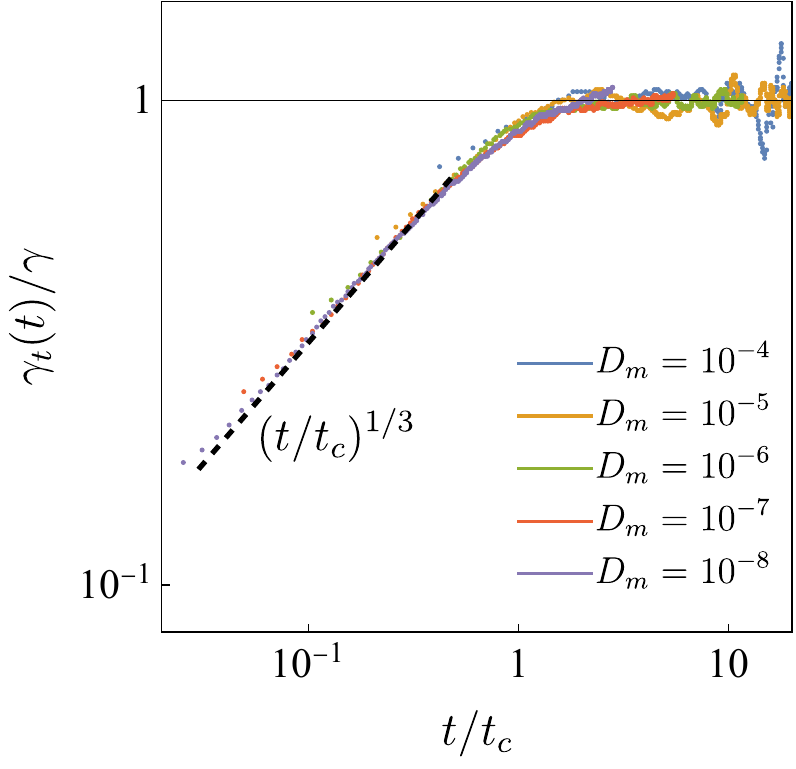}&
\includegraphics[height=0.48\columnwidth]{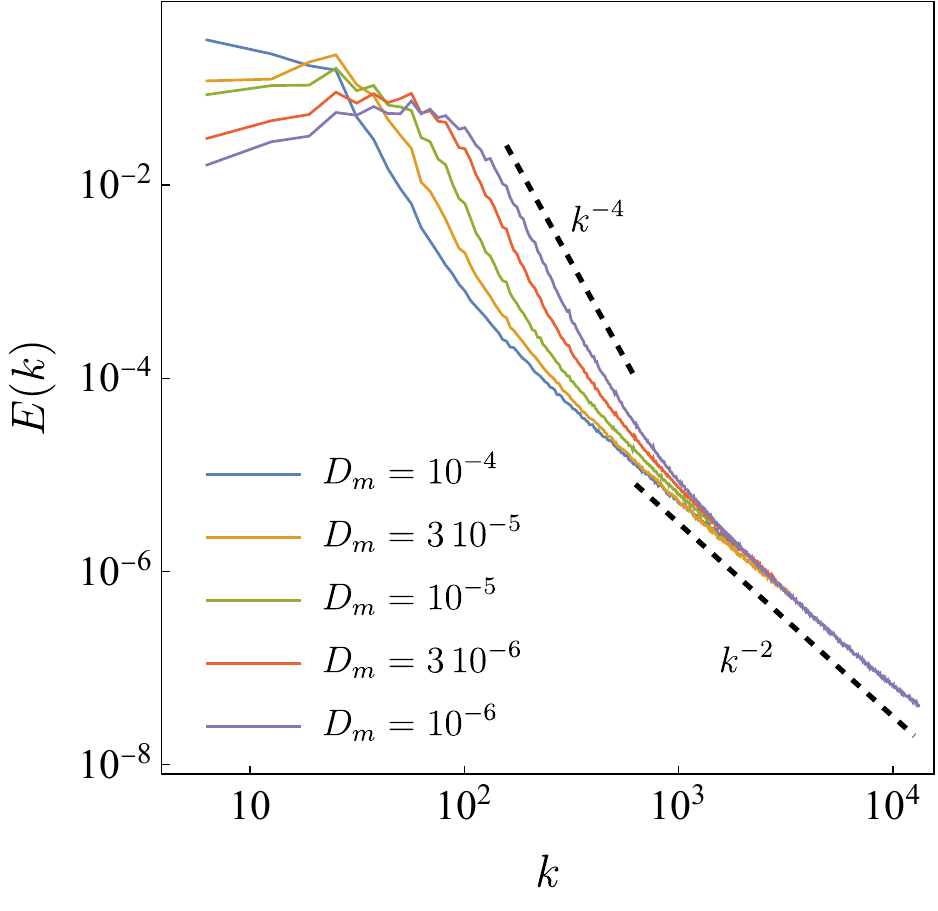}\\
(a) & (b)\\
\includegraphics[height=0.48\columnwidth]{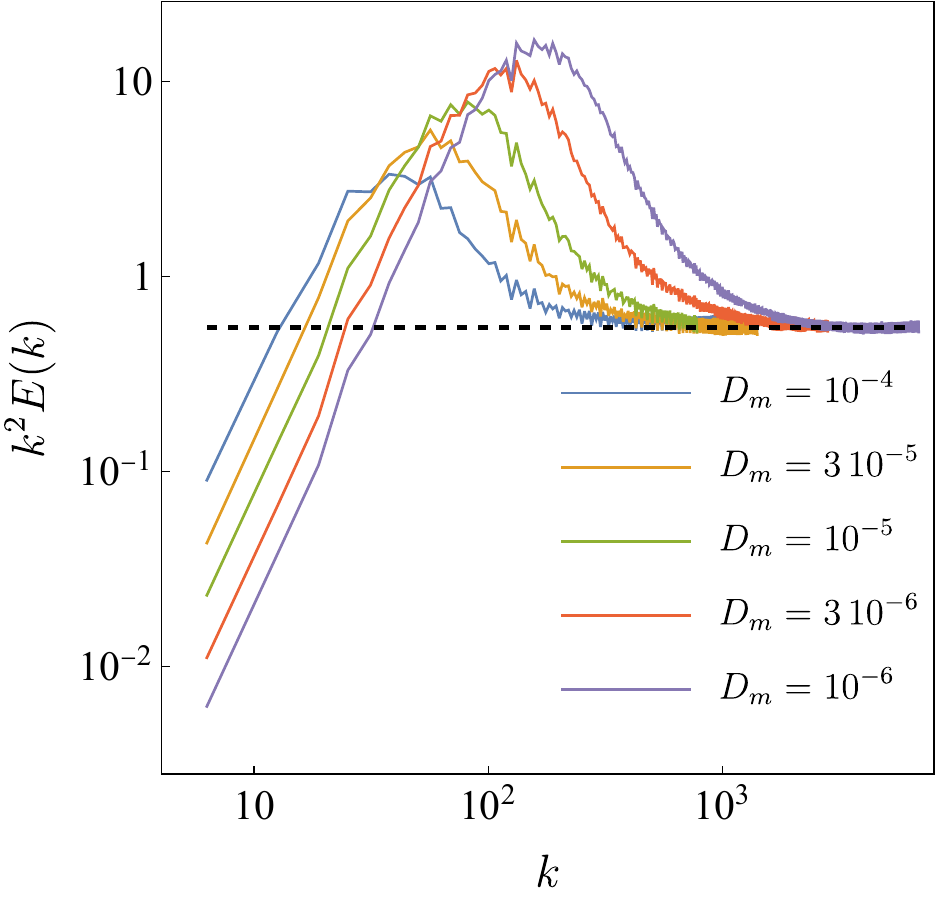}&
\includegraphics[height=0.48\columnwidth]{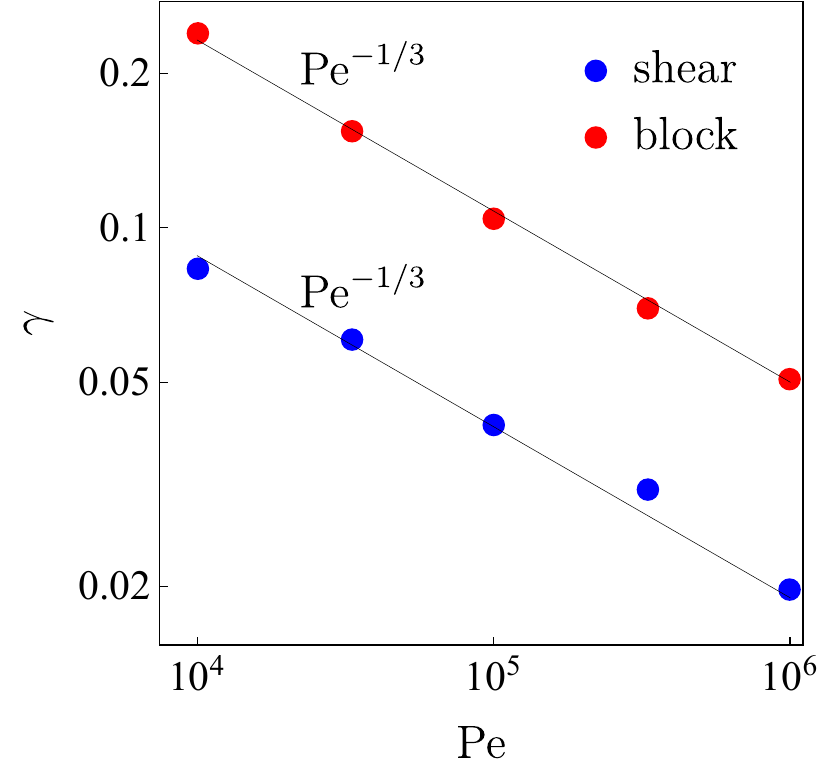}\\
(c) & (d)
\end{tabular}
\caption{Transient and 2D CSD. (a) Transient decay rate $\gamma(t)/\gamma_P$ collapsed on $t/t_c$ with $t_c$ from Eq.~(\ref{eq:tc}); the dashed line is the parameter-free Eq.~(\ref{eq:gammat}). (b) Shell-integrated 2D spectra; the dotted guide is the $k^{-2}$ floor of Eq.~(\ref{eq:asympt}).
(c) Compensated spectra $(Lk)^2E$, showing the $D_m$-independent discontinuity floor. (d) Mixing rate $\gamma\propto D_m^{1/3}$ for PWI families F1 (shear) and F2 (block).}
\label{fig:2d}
\end{figure}

At early times the first set of fronts retains a strong imprint on the system, leading to $E_P\sim k^{-2}$, but for $0<\tau\lesssim 1$, the spectrum develops an Airy core with a surviving $k^{-2}$ range, followed by a transition to the dominant mode, $E_P\rightarrow E_{s,P}$. Before coalescence sets in, every front created since $t=0$ survives, ages are uniform on $[0,t]$, and each front of age $\tau$ dissipates at rate $\sqrt{D_m/2\pi\tau}$, yielding
\begin{equation}
\gamma(t)=\frac{4n_c\lambda}{L}\sqrt{\frac{D_mt}{2\pi}}\qquad(t\ll t_c),
\label{eq:gammat}
\end{equation}
Equation~(\ref{eq:gammat}) holds once the cuts outnumber those of the initial condition, i.e. for $t>\rho_0L/n_c\lambda$ with $\rho_0$ the initial interface density~\cite{SM}. It reproduces the measured rate to $1$--$4\%$, and $\gamma(t)/\gamma_P$ collapses on $t/t_c$ over four decades in $\Pe$ [Fig.~\ref{fig:2d}(a)]. Thus finite-time decay is not a single exponential but a $t^{3/2}$ super-exponential rolling over into a single exponential at $t_c$. This explains why fitting a single stretched exponential $\exp[-(t/\tau)^{\beta}]$ across both regimes~\cite{Wang:2018aa} returns an effective $\beta$ between $1$ and $3/2$ that drifts towards $3/2$ as $\Pe$ increases.

\paragraph*{Multiple Dimensions.} The CSD construction and picture of diffusing and coalescing fronts extends naturally to multiple dimensions. A PWI $\phi$ in any dimension has \emph{codimension-one} discontinuities---points in 1D, lines in 2D, surfaces in 3D---so each front broadens transversely with age $\tau$ as $\sqrt{4D_m\tau}$ and coalesces with its neighbours after $\tau_c\sim s^2/D_m$. For a general PWI the geometry of the discontinuity set determines how Fourier modes are redistributed, and there is no requirement that the resulting spectral operator be isotropic. In particular, a generic $d$-dimensional fractional-Laplacian does not imply a generic $\Pe^{-1/3}$ scaling.

A relevant class is PWIs that generate codimension-one scalar fronts; these can be described in terms of coordinates $(s,r)$ tangent and normal to the front, such that at scales smaller than the radius of curvature of the front, diffusion acts in the normal direction $\partial _t c\approx D_m\partial_r^2 c$. Thus the small-scale front dynamics are locally equivalent to the 1D problem. For a fixed front orientation, with normal $k_n=\mathbf{k}\cdot\mathbf{n}$ and tangential $\mathbf{k}_\bot$ wavevectors, the corresponding transfer operator $T$ has the structure
\begin{equation}
T^{(d)}_{(k_n,\mathbf{p}),(k_n',\mathbf{q})}=T^{(1)}(k_n,k_n')\,\delta_{\mathbf{pq}},
\end{equation}
while diffusion contributes $\exp[-D_m(k_n^2+|\mathbf{k}_\bot|^2)\Delta t]$. Random CS orientations and reconnections couple these sectors but do not introduce a new diffusive length scale. Hence the dominant normal balance therefore retains the 1D coarsening scale: $\bar\ell\sim D_m^{1/3}$, $k_b\sim D_m^{-1/3}$, $\gamma\sim D_m^{1/3}$.

This prediction was tested against two 2D PWI families~\cite{SM}: F1 consisting of randomly positioned strip shear-reconnection events, while F2 randomly partitions the torus into square blocks and applies a block permutation (Fig.~\ref{fig:schematic}). These protocols generate front geometries in fundamentally different ways. The shell-integrated spectrum $E(k)\equiv\sum_{|\mathbf{k}'|\in k\pm\Delta k/2}2|\hat{c}_{\mathbf{k}'}|^2$, with $\Delta k=2\pi/L$ the lattice spacing, for $D_m=10^{-4}-10^{-6}$ under F1 is shown in Fig.~\ref{fig:2d}b (F2 is also similar~\cite{SM}), which exhibits the same $k^{-4}$  and $k^{-2}$ branches as the 1D case, as well as a $D_m$-independent discontinuity floor (Fig.~\ref{fig:2d}c). Hence the high wavenumber spectrum retains the 1D spectrum. Note that the $k^0$ plateau at small $k$ is not observed due to lack of averaging. Fig.~\ref{fig:2d}d also shows that the 2D mixing rates scale as $\gamma\propto D_m^{1/3}$, with different prefactors due to the different PWI protocols.

One important consideration is irreducibility. Similar to 1D CSD~\cite{Kreczak:2017aa}, decoherence is also required in multiple dimensions for the closure (\ref{eq:master}) to $E_p(k,t)$ to hold. For block PWIs, a fixed block lattice does not decohere; random offsets are  required to restore the $D_m^{1/3}$ cascade: this scaling is universal for PWIs which are irreducible on relevant scales.

\paragraph*{Discussion.} As 1D CS is scale-free, random CSD is governed by the coarsening of a population of diffusing fronts. CS creates codimension-one fronts at a rate set by the PWI, while diffusion broadens and annihilates them. Their birth--death balance gives a non-Batchelor coarsening scale $\lb=L(n_c \Pe/2)^{-1/3}$, and algebraic mixing rate $\gamma/\lambda\sim n_c^{2/3}\Pe^{-1/3}$. Hence while mixing is diffusive ($\gamma\sim \Pe^{-1}$) for $\Pe\lesssim 10^3$, at larger $\Pe$, CS accelerates diffusion by $\Pe^{2/3}$ while remaining asymptotically diffusion-limited; $\gamma/\lambda\rightarrow 0$ as $\Pe\rightarrow\infty$.

The same mechanism extends to transient and 2D mixing. Transient simulations using two geometrically distinct, measure-preserving PWIs recover $\gamma\propto D_m^{1/3}$ and the characteristic $k^{-4}\to k^{-2}$ spectral structure. This suggests that the $1/3$ exponent is a geometric consequence of irreducible renewal of codimension-one fronts.

These results place discontinuous mixing in the non-equilibrium universality class of diffusion-controlled coarsening~\cite{Racz:1985aa,Krapivsky:2010aa}, so that the extensive exact results available for that class~\cite{Torney:1983aa,ben-Avraham:1990aa} may be brought to bear on higher-order mixing statistics and on reactive and multicomponent transport. This framework applies wherever CS organise transport, from
granular~\cite{Juarez:2010aa,Juarez:2012aa,Christov:2010aa,Zaman:2018aa,Park:2016aa}, porous~\cite{Lester:2025ae} and fractured~\cite{Ascione:2026aa} media to biological tissues, valved systems~\cite{Smith:2016aa} and microfluidic devices~\cite{Smith:2018aa,Smith:2019aa}.

\paragraph*{Data availability statement.} The data that support the findings of this article are publicly available~\cite{DB}.

\begin{acknowledgments}
\end{acknowledgments}

\end{document}